# Time-synthetic optical neural networks with stable programmable gain

Bei Wu, Yudong Ren, Rui Zhao, Haiyao Luo, Fujia Chen, Li Zhang, Lu Zhang, Hongsheng Chen* and Yihao Yang*

*[1] State Key Laboratory of Extreme Photonics and Instrumentation, ZJU-Hangzhou Global Scientific and Technological Innovation Center, Zhejiang University, Hangzhou 310027, China.*
*[2] International Joint Innovation Center, The Electromagnetics Academy at Zhejiang University, Zhejiang University, Haining 314400, China.*
*[3] Key Lab. of Advanced Micro/Nano Electronic Devices & Smart Systems of Zhejiang, Jinhua Institute of Zhejiang University, Zhejiang University, Jinhua 321099, China.*
*[4] Shaoxing Institute of Zhejiang University, Zhejiang University, Shaoxing 312000, China.*

** Corresponding authors: hansomchen@zju.edu.cn (H. Chen); yangyihao@zju.edu.cn (Y. Yang)*

**Abstract**

Optical neural networks (ONNs) offer ultrafast and energy-efficient artificial intelligence, yet their effective depth remains fundamentally limited because the core linear transformations are overwhelmingly passive, and cumulative loss rapidly degrades the signal-to-noise ratio. Introducing optical gain into spatial photonic meshes could, in principle, counteract this decay, but such amplification is notoriously unstable owing to unavoidable feedback paths and parasitic reflections. Here, we overcome this long-standing limitation by integrating programmable gain into a time-synthetic ONN, where computation unfolds through strictly forward temporal evolution rather than spatial interferometric layers. This causal topology suppresses the backward channels that trigger gain-induced instabilities, enabling stable loss compensation and substantially extending the network's usable depth. Numerical analysis and in-situ experiments with more than 30,000 effective gates demonstrate robust performance on image-classification tasks, establishing gain-assisted time-synthetic ONNs as a stable, scalable, and programmable pathway toward deep photonic intelligence beyond the limitations of predominantly passive architectures.

## Introduction

Optical neural networks (ONNs) have attracted growing interest as a promising platform for accelerating artificial intelligence (AI), owing to the massive parallelism, broad bandwidth, and negligible Joule heating offered by photons. Over the past decade, interferometric photonic processors[1-4], diffractive optical networks[5,6], and metasurface-based architectures[7-9] have shown great performance in low-power inference and optical matrix multiplication. Nevertheless, despite active elements are sometimes used, like in optical-electrical-optical conversion stages or nonlinear activation modules, the computational cores of current ONNs are predominantly passive[10]. Since these passive transformations are either energy-conserving or dissipative, they can only achieve unitary or contraction matrices[11]. As optical signals propagate through such networks, losses caused by beam splitters, modulators, waveguides, and coupling interfaces unavoidably accumulate[12]. In deep circuits, the cumulative losses quickly lower the signal-to-noise ratio (SNR), making outputs dominated by thermal and detector noise and preventing current ONNs from reaching the depths required for modern AI workloads.

Incorporating optical gain offers a natural route to overcoming these limitations. In principle, controlled amplification could counteract propagation losses, preserve signal strength across long optical paths, and expand the transformation space from unitary operations to any complex-valued matrices[13]. Yet integrating gain directly into the computational pathways of an ONN is far from straightforward. Even small amplification imbalances can cause runaway power growth, chaotic oscillations, or mode-selective instabilities—phenomena that are greatly amplified in spatial photonic meshes dense with unavoidable feedback loops and unwanted parasitic reflections[14]. Consequently, the majority of ONN architectures choose to avoid inline gain, accepting the resulting performance degradation as an inevitable compromise. This tension has caused a long-standing dilemma in this field: while gain is essential for maintaining computational fidelity in deep optical networks, its direct incorporation tends to destabilize the very systems it is meant to enhance.

In this work, we address this contradiction by developing a fully programmable ONN that integrates gain within a time-synthetic dimension instead of a spatial photonic mesh[15-18]. A pair of coupled optical loops with slightly different lengths creates a temporal lattice in which optical pulses evolve through discrete time steps while experiencing dynamically programmed gain, loss, and phase shifts. Because computation only happens in the forward direction of time[19], the architecture is inherently causal: pulses never retrace their paths, and energy flow does not go backwards. This temporal causality stops the feedback pathways that usually cause gain-induced instabilities in spatial systems. As a result, the network can reliably harness amplification to counteract losses and sustain deep computation, as long as the programmed gain stays below the point where instability would happen.

The time-synthetic dimension also provides a powerful scalability advantage. Network depth is based on the number of round trips, not the number of physical components, enabling one compact unit to emulate tens of thousands of effective optical gates[20,21]. This reduces the $O(N^2)$ scaling of spatial photonic processors into an $O(1)$ footprint and makes the network more robust. To enable the network learning in realistic hardware, we further develop an in-situ optical training scheme that gets gradients directly from measured intensities, allowing the system to adapt to hardware imperfections, thermal drift, and accumulated noise[22-24]. Numerical studies indicate that gain maintains signal strength and stabilizes gradient propagation in deep temporal networks, while experiments on a coupled-loop platform with more than 30,000 effective time gates exhibit robust inference on datasets such as MNIST and CIFAR-10. These results collectively demonstrate that the incorporation of gain into a time-synthetic dimension offers a stable, scalable, and programmable pathway to deep optical neural networks, effectively addressing the depth and fidelity constraints associated with passive photonic processors.

## Results

### Gain-assisted optical computing in time-synthetic dimension

The temporal optical computing architecture is based on a time-synthetic lattice, which has been widely used to explore the exotic non-Hermitian or topological physics[25-30]. Specifically, optical pulses travel through two optical loops of slightly different lengths, coupled via a variable beam splitter (BS), as shown in Fig. 1a. A 1550 nm pulsed laser injects an optical pulse into the longer loop, while the shorter loop integrates a programmable Mach-Zehnder modulator (MZM) and a phase modulator (PM), to dynamically control the gain/loss and phase shifters of optical pulses. The optical power in each loop is monitored by photodetectors in real time; see Methods for experimental details.

The length disparity between the loops introduces a temporal offset: Pulses in the shorter loop advance by $\Delta t$ per round trip, while those in the longer loop accumulate an equivalent delay. This temporal offset $2\Delta t$ per cycle emulates spatial displacement in a synthetic dimension, with the round-trip count defining the time layer. Consequently, pulse propagation manifests as trajectories through a two-dimensional (2D) effective spacetime mesh lattice, which underpins our ONN architecture (Fig. 1b).

Pulse dynamics in the time-synthetic lattice obey the modified discrete quantum walk equations[31],

$$u_m^{n+1} = G[\cos(\beta)\, u_{m+1}^n + i \sin(\beta)\, v_{m+1}^n] e^{i\varphi} \tag{1}$$

$$v_m^{n+1} = i \sin(\beta)\, u_{m-1}^n + \cos(\beta)\, v_{m-1}^n \tag{2}$$

where $u_m^n/(v_m^n)$ represents the complex amplitude of the optical pulse at lattice position $m$ and time layer $n$ in the shorter/(longer) loop. The parameter $\beta = \beta(m, n)$ represents the beam-splitting ratio, $G = G(m, n) \in [e^{-0.3}, e^{0.3}]$ represents the gain/loss, and $\varphi = \varphi(m, n) \in [0, 2\pi)$ represents the phase shifter. Unlike the SU(2) gates in conventional photonic circuits that are unitary owing to the system's passivity, the time gates involve programmable gain and loss (Fig. 1c), which are inherently non-Hermitian. Compared with the unitary transformations confined to the Bloch sphere's surface[11], the non-Hermitian operators occupy the entire, theoretically unbounded volume between the limits of gain and loss (Fig. 1d). The phase shifter enables tunable *z*-axis rotations, and the variable BS governs *x*-axis rotations, as shown in Fig. 1e. Even when the beam-splitting ratio is fixed, the phase difference between two cascaded time gates enables arbitrary *x*-axis rotations. The non-Hermitian property enables the time gate to implement arbitrary complex-valued transformations and compensates for optical loss, which significantly enhances the depth and expressivity of ONNs.

**Time-synthetic ONN architecture**

The standard ANNs consist of a set of input artificial neurons connected to hidden layers and an output layer. In our time-synthetic ONN architecture, each lattice site acts as an artificial neuron, with its complex amplitude representing the neuron's activation value. Consequently, the network depth is defined by the round trips, and the width by the number of optical pulses in the loops. The output can be expressed as $\widehat{Y^{out}} = W(G^n, \varphi^n) \ldots W(G^1, \varphi^1) W(G^0, \varphi^0) \psi_i$, where $\psi_i$ represents the injected optical pulse. The programmable feature of weight matrix $W(G^n, \varphi^n)$, hereafter abbreviated as $W^n$, is optically realized via the tunable gain/loss $G^n = [G_1^n, \ldots, G_M^n]$ and phase shifters $\varphi^n = [\varphi_1^n, \ldots, \varphi_M^n]$, respectively, with the specific expression given in Methods. Here, the superscript $n$ in the variables $G_m^n$ and $\varphi_m^n$ represents the time layer, and the subscript $m$ represents the position.

The nonlinear activation function in this linear system is achieved through structural nonlinearity[9,32]. While the input-output relationship governed by the propagating operator, $W^n$, is linear, the dependence of the propagating operator itself on system parameters introduces nonlinearity. Specifically, our architecture encodes input signals onto either the phase shifters $\varphi_x$ or the gain/loss factors $G_x$ at designated lattice sites, while the remaining parameters ($G_\theta$, $\varphi_\theta$) serve as learnable weights. Encoding the input signals into the phase shifters yields a complex exponential nonlinearity: $\widehat{Y_i^{out}} = f_{NL}(\varphi_x) = \alpha_{m,n}(G_\theta, \varphi_\theta)e^{i\varphi_x}$, where $\alpha_{m,n}$ is a function of learnable weights $G_\theta$ and $\varphi_\theta$. Alternatively, encoding the input signals into the gain/loss factors, with an M-fold duplication (M = 2 in this work), produces a polynomial nonlinearity: $\widehat{Y_i^{out}} = f_{NL}(G_x) = \alpha_{m,n}(G_\theta, \varphi_\theta)G_x^{\ M}$. Beyond the structural nonlinearity employed in this work, nonlinear responses can be achieved by harnessing the intrinsic nonlinear characteristics of photonic components, which include, but are not limited to: optoelectronic feedforward[33], Kerr self-phase modulation in silica[34], and periodically poled lithium niobate waveguides[35].

During training, the mean-square error (MSE) is employed to quantify the deviations between the predicted and target outputs: $\mathcal{L} = (\widehat{Y^{out}} - Y)^\dagger(\widehat{Y^{out}} - Y)/2$, where the target outputs are one-hot encoded. The differentiable nature of quantum walk dynamics enables error backpropagation, which is a cornerstone technique for training ANNs. This technique propagates error signals backward through the network to compute gradients of learnable weights via the chain rule.

**Theory of the gain-assisted ONNs**

To validate the computational power of the time-synthetic ONN, we benchmark its performance on the MNIST handwritten digit classification task. The network propagates through 40 round trips, where input images (normalized to $[0, \pi]$) are encoded onto the phase shifters of optical pulses in time layers $L_{12}$ - $L_{17}$, $L_{20}$ - $L_{21}$, and $L_{24}$ - $L_{25}$. The network's learnable weights are physically realized by using phase shifters to control their sign (via constructive/destructive interference) and optical gain/loss factors to set their amplitude (Fig. 2a). This lets the network enhance important features and hide unimportant ones. In the end, it sends the optical pulses to one of ten output positions in the shorter loop, each of which represents a digit class, as shown by the optical propagation dynamics in Fig. 2b.

Optical gain is pivotal for counteracting propagation loss, a fundamental challenge in deep photonic circuits. Without amplification, the optical signal decays exponentially with network depth and can be obscured by the photodetector's intrinsic electronic noise, which leads to a catastrophic drop in the SNR and, consequently, the classification accuracy (Fig. 2c). In this case, the SNR is given by $SNR = I_s^2/(\langle i_{thermal}^2\rangle+\langle i_{shot}^2\rangle)$. Here, $I_s$ is the signal photocurrent. The mean-square thermal noise current[36] is $\langle i_{thermal}^2\rangle = 4kTB_e/R$, where $k$ is the Boltzmann constant, $T$ is the absolute temperature, $B_e$ is the photodetector's electronic bandwidth, and $R$ is the load resistance. The mean-square shot noise current[37] is $\langle i_{shot}^2\rangle = 2eI_sB_e$, where $e$ is the elementary charge.

However, optical gain also introduces amplified spontaneous emission (ASE) noise, with an optical power[38] of $P_{ASE} = 2n_{sp}(G - 1)h\nu B_0$, where $n_{sp}$ is the spontaneous emission factor, $G$ is the gain factor, $h$ is Planck's constant, $\nu$ is the optical frequency, and $B_0$ is the optical bandwidth of the amplifier. This ASE noise, in turn, generates a signal-spontaneous beat noise current at the photodetector, with a mean-square value of $\langle i_{sig\text{-}sp}^2\rangle = 2I_sI_{ASE}B_e/B_0$. Consequently, the SNR equation is modified to $SNR = I_s^2/(\langle i_{thermal}^2\rangle+\langle i_{shot}^2\rangle+\langle i_{sig\text{-}sp}^2\rangle)$. As depicted in Fig. 2d, the SNR initially increases sharply with optical gain as the amplified signal power overcomes the photodetector's electronic noise floor, before saturating once the ASE noise becomes dominant. In this saturation regime, the SNR is sufficiently high to preserve the computational fidelity.

To validate the pivotal role of optical gain, we evaluate the network's performance under two distinct conditions: with and without optical gain. In both scenarios, the input optical power is 1 µW, and the system is subjected to a root-mean-square (RMS) thermal noise current of 1.28 µA, corresponding to a 5 GHz electronic bandwidth and a 50 Ω load resistance at room temperature. In the gain-assisted scenario, the cumulative optical gain is 15 dB, and the time-synthetic ONN achieves a test accuracy of 97%, characterized by a near-diagonal confusion matrix (Fig. 2e) and a clearly separable feature space in the t-SNE visualization (Fig. 2f). The output intensity distributions for representative test samples in Fig. 2b remain robust against noise, ensuring correct classification as shown in Fig. 2g. In stark contrast, without optical gain, the network's classification accuracy plummets to 55.3% (Fig. 2h), and the learned feature space becomes indistinguishable (Fig. 2i). It leads to catastrophic misclassifications, with a half of representative test samples being incorrectly identified (Fig. 2j). This demonstrates that the optical gain is not merely a compensatory mechanism but an enabling technology, critical for scaling ONNs to the sophisticated computational domains such as natural language processing and multimodal tasks.

**In-situ training of a large-scale time-synthetic ONN**

While numerical simulations demonstrate exceptional accuracy, experimental deployment with pre-trained parameters often exhibits performance degradation, mainly due to system imperfections and calibration drift. To address these challenges, we implement a two-stage training protocol. In the first stage, the time-synthetic ONN is pre-trained in-silico, learning the ideal propagation dynamics from simulated data[1]. Second, it is fine-tuned using an in-situ optical training framework (Fig. 3a) to adapt to real-world hardware noise and imperfections[22-24]. Unlike in-silico training, which relies on an ideal computational model, in-situ training derives gradients directly from experimental measurements, significantly improving the accuracy.

To overcome the phase-blindness of photodetectors, the in-situ training framework forgoes phase modulation to reconstruct complex amplitudes from detected optical intensities. The D-dimensional gain/loss gradients are derived via the chain rule[22]: $\frac{\partial \mathcal{L}}{\partial G_\theta^n} = \mathcal{R}\left\{(\widehat{Y^{out}} - Y)^\dagger \frac{\partial W^n}{\partial G_\theta^n} \widehat{Y^n}\right\}$ for the output layer, and $\frac{\partial \mathcal{L}}{\partial G_\theta^{n-1}} = \mathcal{R}\left\{\frac{\partial \mathcal{L}}{\partial \widehat{Y^n}} \frac{\partial W^{n-1}}{\partial G_\theta^{n-1}} \widehat{Y^{n-1}}\right\}$ for hidden layers, where $\mathcal{R}\{\cdot\}$ gives the real part. Error propagation follows: $\frac{\partial \mathcal{L}}{\partial \widehat{Y^n}} = (W^n)^{\mathrm{T}}(\widehat{Y^{out}} - Y)^\dagger$ for the output layer, and $\frac{\partial \mathcal{L}}{\partial \widehat{Y^{n-1}}} = (W^{n-1})^{\mathrm{T}} \frac{\partial \mathcal{L}}{\partial \widehat{Y^n}}$ for hidden layers. The items $\psi_i$, $\widehat{Y^n}$, and $W^n$ are measurable; $\frac{\partial W^n}{\partial G_\theta^n}$ is constant; and $Y$ is given, as detailed in Methods. Subsequently, gain/loss factors are updated via gradient descent: $G_\theta \leftarrow G_\theta - \eta \frac{\partial \mathcal{L}}{\partial G_\theta}$, where $\eta$ is the learning rate.

While this work does not measure the optical phase, existing methods such as residual carrier modulation[40] merit consideration. Additionally, several in-situ training frameworks eliminate the need for phase recovery, including the forward-forward algorithm[28] and the stochastic perturbation method[29].

**Experimental demonstration of object recognition in the time-synthetic ONN**

We experimentally evaluate the time-synthetic ONN in a coupled optical-loop system, as shown in Fig. 3b. The network propagates through 40 round trips and is tested on the CIFAR-10 benchmark dataset. Input RGB images are converted to grayscale, duplicated, normalized to $[e^{-0.3}, e^{0.3}]$, and encoded onto gain/loss factors in designated time layers. The other gain/loss factors serve as learnable weights to direct the optical pulses toward the target output positions in the shorter loop.

To visualize the in-situ training dynamics, we track the learning process for a representative test sample. As shown in Fig. 3c, the time-synthetic ONN progressively steers the optical pulses from the initial random state toward the target output. This successful learning trajectory is a direct result of optical gain, which mitigates the vanishing gradient problem during training. Figure 3d shows that at the beginning of training, the randomly set gain/loss factors give large gradients, which means there is a strong error signal. The gradients slowly get smaller as the network gets closer to the best solution, signifying a stable optimization process.

To test the network's robustness, we add Gaussian noise to the input signals, as shown in Fig. 3e. Remarkably, the classification accuracy remains stable even as the noise standard deviation approaches 0.3, demonstrating exceptional error resilience with in-situ training strategy. For further analysis, Fig. 4a tracks the output intensity distributions for more samples during the experiment, which converge to target outputs rapidly. Although transient mechanical noise occasionally induces oscillations between error states, in-situ training enables prompt recovery.

After training with the AdaGrad optimizer, the gain/loss factors converge to the patterns shown in Fig. 4b, enabling the time-synthetic ONN to achieve a test accuracy of 86.5% (Fig. 4c). The accuracy is mainly affected by two factors: high similarity across data samples, which causes the network outputs to have overlapping feature spaces (Fig. 4d), and the calibration drift of the optical tunable filter (OTF). The in-situ training framework can compensate for the calibration drift to some extent, and we evaluate its effectiveness by performing 10 × 10 matrix operations. The matrix fidelity is defined as $F = (1 - \frac{|\hat{T} - T|}{\hat{T} + T}) \times 100\%$, where $\hat{T}$ and $T$ represent the achieved and target matrices, respectively. As shown in Fig. 4e, while in-silico training only achieves a median fidelity of 94.8%, in-situ training improves it to 98.5%. We further visualize the optical pulse propagation dynamics for representative test samples (Fig. 4f), which confirms that the optical pulses are precisely directed to the target positions.

The time-synthetic ONN achieves high scalability by orchestrating pulse evolution with nanosecond precision. Through a precisely engineered 50 ns pulse width and 196 ns temporal separation between the shorter and longer loops, the system can sustain about 251 optical pulses in the 5 km coupled optical loops. The optical pulses exhibit high signal fidelity after propagating through 124 round trips, generating 31,124 effective gates that exceed the state-of-the-art programmable photonic circuits[41,42].

The performance can be further enhanced via two key strategies: miniaturized integration and enhanced computational capability. First, reducing the length of optical loops is critical for integration. Specifically, the minimum loop length, $L$, is governed by the relation $L \geqslant pc\Delta t/n_{eff}$, where $p$ is the number of optical pulses in the loop, $c$ is the speed of light in vacuum, $\Delta t$ is the temporal separation between adjacent pulses, and $n_{eff}$ is the effective refractive index of the waveguide. Thus, the loop length can be shortened by decreasing $\Delta t$, which is typically constrained by the modulation speed of MZM or PM. Beyond hardware improvements, a fully loaded pulse propagation scheme (detailed in Supplementary Note 1) can be adopted to maximize the utilization of the computational space in a given loop length. To elaborate, we provide a detailed integrated photonic circuit design in Supplementary Note 3, yielding a theoretical area efficiency of 13.6 Tera-FLOPs/mm$^2$/s (equivalent to 6.8 Tera-MACS/mm$^2$). Second, to significantly boost computational capability, further improvements involve leveraging mature parallelization technologies from optical communications, including wavelength division multiplexing and multimode waveguides[43].

**Discussion**

In this work, we resolve a long-standing contradiction in ONNs by showing that, although optical gain is essential for maintaining signal fidelity and achieving deep computation, its instability in spatial photonic meshes arises from feedback and parasitic pathways rather than from amplification itself. By relocating the network into a time-synthetic dimension, where computation proceeds strictly forward in time, gain becomes a stable and programmable resource that allows for controlled non-Hermitian transformations, effective loss compensation, and a much deeper network. Combined with in-situ optical training and a coupled-loop platform that supports more than 30,000 effective gates, our results show that gain-assisted time-synthetic ONNs can break through the depth, fidelity, and scalability limits of passive architectures and establish temporal synthetic dimensions as a powerful way to stabilize active photonic computing.

The current architecture prioritizes stability and compactness, but its throughput is limited because it only has one modulator and processes data in order. This constraint can be alleviated by using the natural parallelism of photonics. Spectral multiplexing, spatial-mode multiplexing, or multimode waveguide platforms could greatly increase the number of computational channels, while sub-picosecond pulse engineering could also speed up the number of operations per round trip. Integration onto photonic chips would make them more compact, and combining temporal synthetic dimensions with spatial architectures could create hybrid ONNs that keep the scalability benefits of synthetic dimensions while getting the high throughput of spatial processors[44]. These directions illustrate that the gain-stabilized time-synthetic paradigm is compatible with existing photonic technologies and is also well-suited to support the next generation of scalable, high-performance photonic AI systems.

**Acknowledgments**
The work at Zhejiang University sponsored by the Key Research and Development Program of the Ministry of Science and Technology under Grants No.2022YFA1405200 (Y.Y.), No.2022YFA1404900 (Y.Y., H.C.), and No.2022YFA1404704 (H.C.), the National Natural Science Foundation of China (NNSFC) under Grants No. 62175215 (Y.Y.), and No.61975176 (H.C.), the Key Research and Development Program of Zhejiang Province under Grant No.2022C01036 (H.C.), the Fundamental Research Funds for the Central Universities (2021FZZX001-19) (Y.Y.), and the Excellent Young Scientists Fund Program (Overseas) of China (Y.Y.).

**Author contributions**
Y.Y. conceived the idea of this research. B.W. proposed algorithm structure and conducted the experiment. R.Z., L.Z., Y.R., F.C and assisted in the experimental setup. H.L. assisted in the intelligent algorithm. Y.Y. and B.W. wrote the paper. All authors shared their insights and contributed to discussions on the results. H.C. and Y.Y. supervised the project.

**Data availability**
The data that support the findings of this study are available from the authors on reasonable request.

**Competing interests**
The authors declare no competing financial interests.

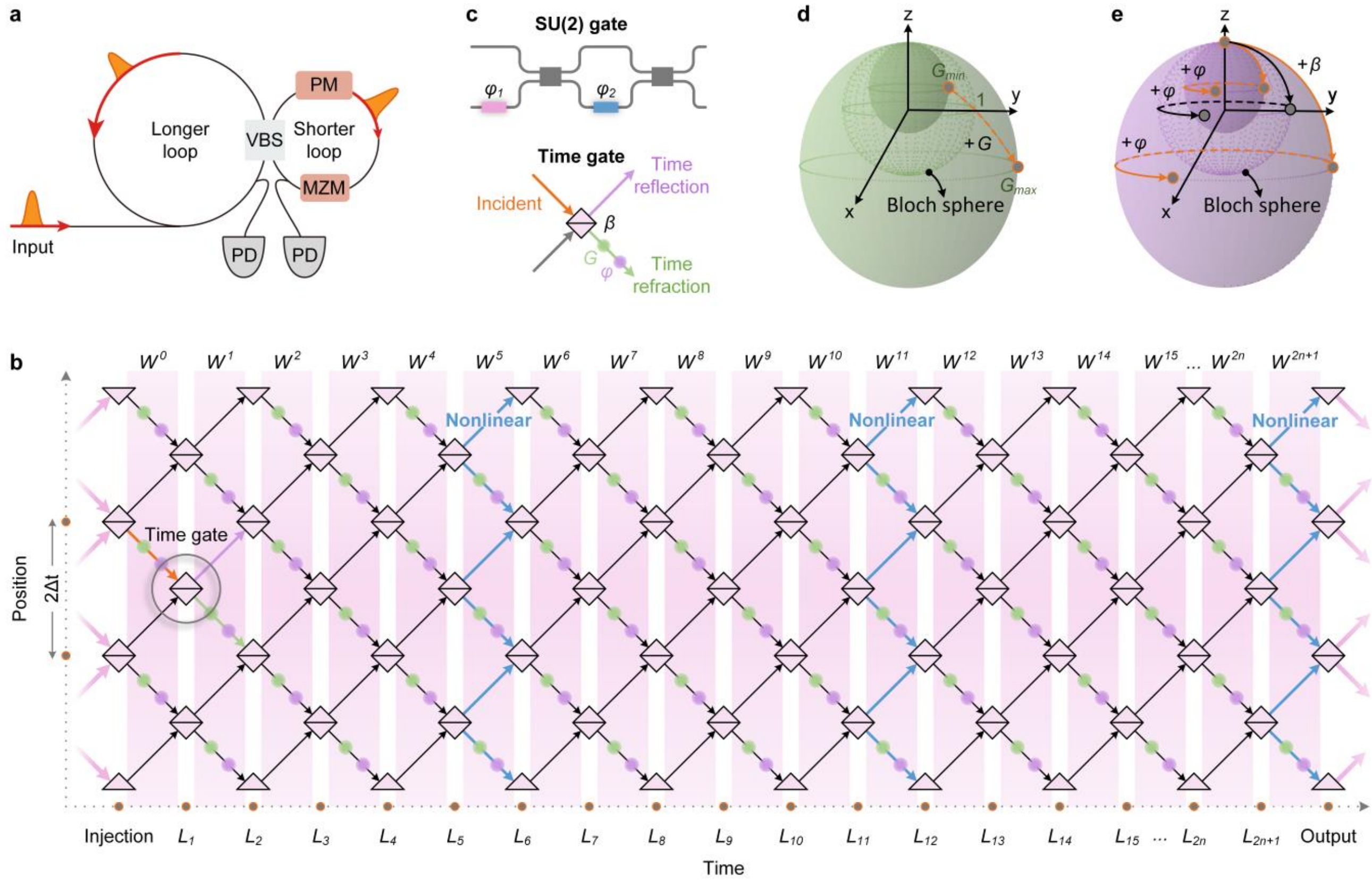

**Fig. 1. Architecture of the gain-programmable ONNs in a time-synthetic dimension. a** Coupled optical loops. A 1550 nm pulsed laser injects an optical pulse into the longer loop, while the shorter loop integrates an MZM and a PM to control the gain/loss and phase shifters of optical pulses, respectively. As the pulses propagate in the system, the optical power in each loop is monitored by photodetectors. **b** Time-synthetic ONN mapped from the pulse propagation in **a**. The pink diamonds represent BSs, while the green and purple circles depict gain/loss and phase shifters, respectively. The blue layers represent nonlinear activation functions. **c** Top: SU(2) gate. Blue and pink boxes represent the phase shifters $\varphi_1$ and $\varphi_2$, respectively. Bottom: non-Hermitian time gate. Green and purple circles represent the gain/loss $G$ and phase shifter $\varphi$, respectively. **d** Scaling operator for the gain/loss modulation of the time gate. This scaling mechanism enables non-Hermitian operators to evolve in the volume bounded by gain and loss limits. **e** Rotation operator for the phase shifter and the variable BS of the time gate. The phase shifter enables tunable *z*-axis rotations, while the variable BS governs *x*-axis rotations. The unitary transformation implemented by the SU(2) gate corresponds to a rotation on the surface of the Bloch sphere, as depicted by the black line.

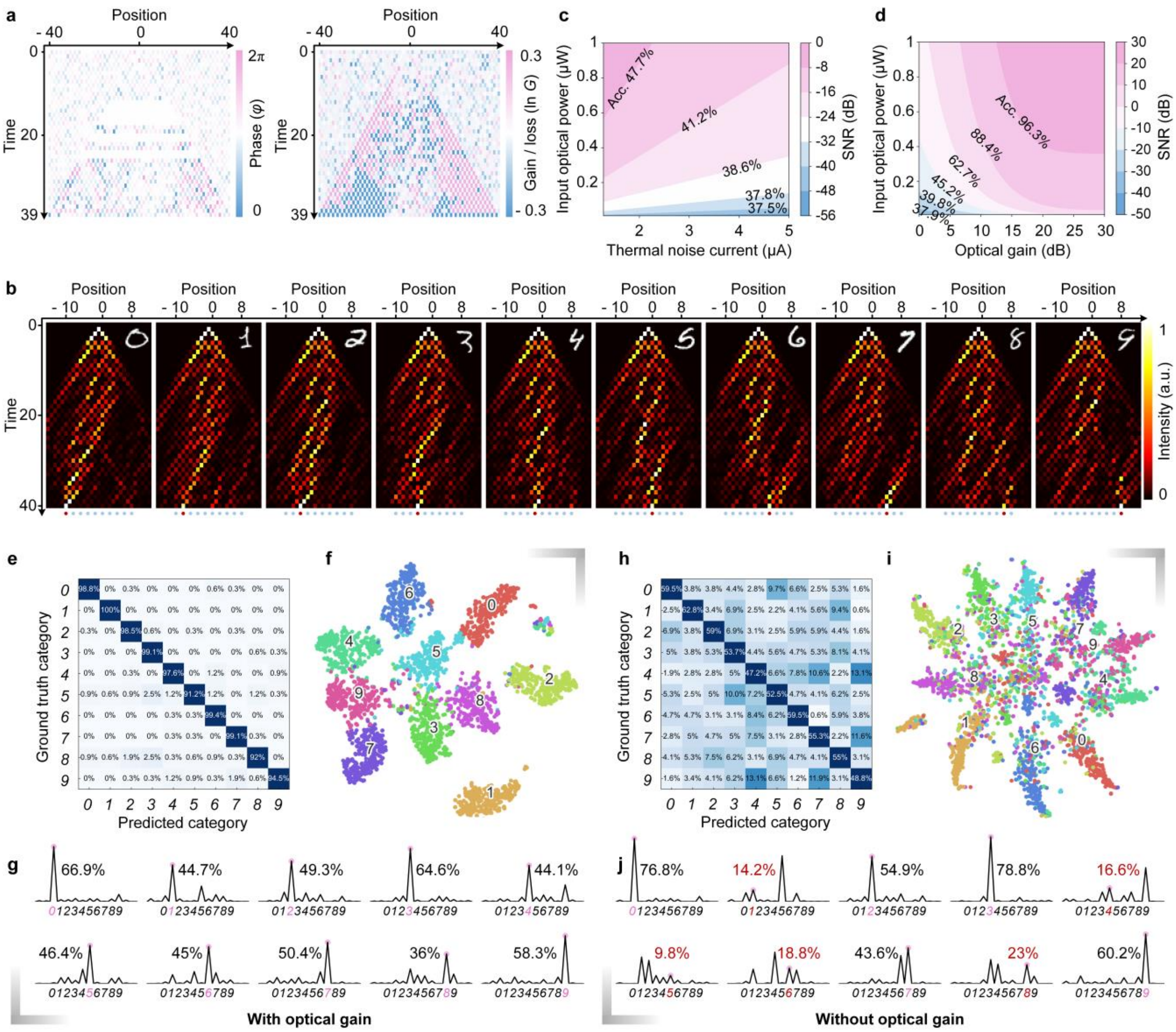

**Fig. 2. Theory of the gain-assisted ONNs. a** Optimized phase shifters (left) and gain/loss factors (right). Input images are encoded onto the phase shifters in time layers $L_{12}$ - $L_{17}$, $L_{20}$ - $L_{21}$, and $L_{24}$ - $L_{25}$ (blank regions). **b** Optical pulse propagation in the shorter loop for ten representative digits. **c** SNR as a function of input optical power and thermal noise, with corresponding test accuracies annotated. **d** SNR as a function of input optical power and gain/loss factors, with an RMS thermal noise current of 1.28 μA. **e**-**g** Network performance in the gain-assisted scenario, demonstrated by (**e**) the confusion matrix with an accuracy of 97%, (**f**) a clearly separable feature space in the t-SNE visualization, and (**g**) robust output intensity distributions. **h**-**j** Collapse of computational performance without optical gain, demonstrated by (**h**) a drop in accuracy of 55.3%, (**i**) an indistinguishable feature space, and (**j**) corrupted output intensities leading to misclassification.

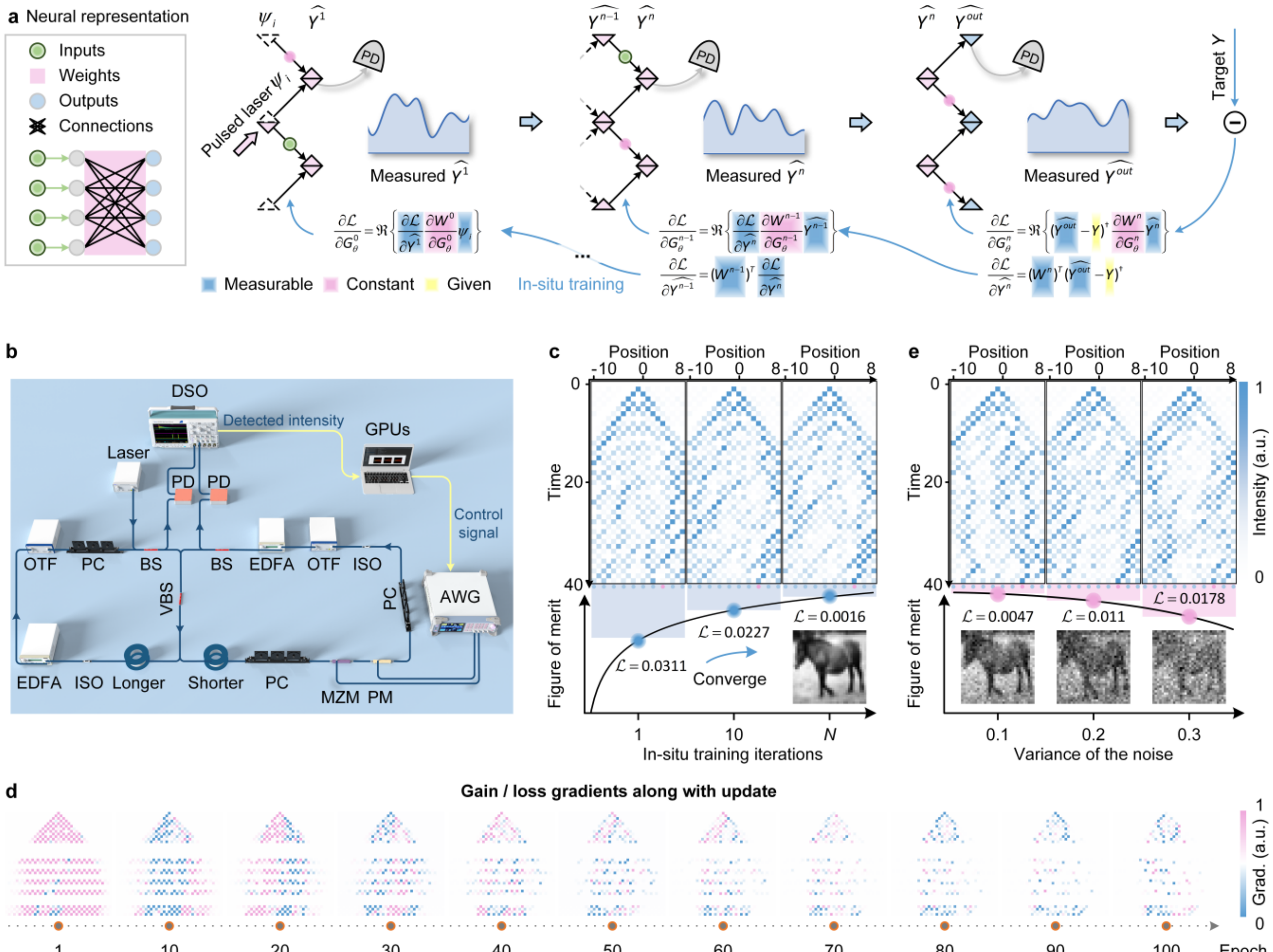

**Fig. 3. In-situ training of a large-scale time-synthetic ONN.** **a** Schematic of the in-situ training framework. The items $\psi_i$, $\widehat{Y^n}$, and $W^n$ are measurable; $\frac{\partial W^n}{\partial G_\theta^n}$ is constant; and $Y$ is given. **b** Simplified coupled optical-loop setup. **c** Optical pulse propagation in the shorter loop for a representative test sample during training. **d** Evolution of normalized gain/loss gradients during training. **e** Experimental noise analysis. The input signal is intentionally corrupted with varying levels of Gaussian noise to test the network's robustness. DSO: digital storage oscilloscope; GPU: graphic processing unit; PD: photodetector; OTF: optical tunable filter; PC: polarization controller; BS: beam splitter; VBS: variable beam splitter; ISO: isolator; EDFA: erbium-doped fiber amplifier; MZM: Mach-Zehnder modulator; PM: phase modulator; AWG: arbitrary waveform generator.

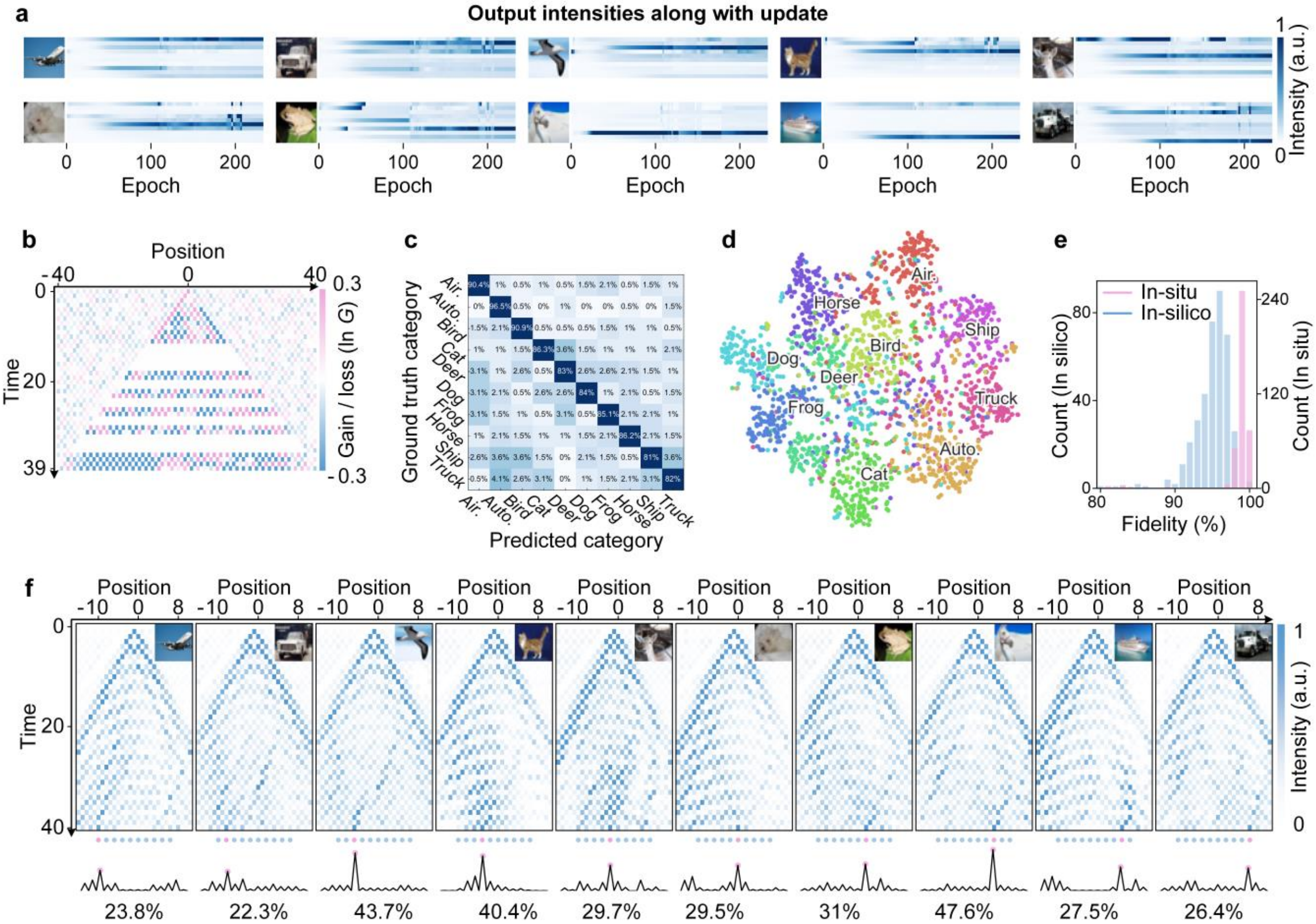

**Fig. 4. Experimental demonstration of object recognition in the time-synthetic ONN. a** Evolution of output intensities for representative test samples during training. **b** Optimized gain/loss factors. Input signals are encoded onto the gain/loss factors in the time layers $L_{12}$ - $L_{17}$, $L_{20}$ - $L_{21}$, $L_{24}$ - $L_{25}$, $L_{28}$ - $L_{29}$, and $L_{32}$ - $L_{35}$ (blank regions). **c** Confusion matrix for test images, with an accuracy of 86.5%. **d** t-SNE dimensionality reduction and visualization of the output features. **e** Fidelity of 10 × 10 matrix operations. The pink/blue bars represent the matrix fidelity achieved through in-situ/silico training, revealing that in-situ training enhances the median fidelity from 94.8% to 98.5%. **f** Optical pulse propagation in the shorter loop (top) and the normalized output intensities (bottom) for ten representative objects.

## Methods

### Experimental details

Our experimental architecture employs dual optical loops coupled via a BS, as depicted in Fig. 3b and Extended Data Fig. 1. A 1550 nm optical signal generated by a distributed feedback laser undergoes pulse shaping via an acousto-optic modulator (AOM), producing a square pulse with >50 dB extinction ratio and 50 ns temporal width, which is injected into the longer loop. The variable BS subsequently splits each pulse according to a predefined beam-splitting ratio ($\beta = \pi/4$ in this work), then directs the split pulse into the longer and shorter loops, respectively. Real-time control of the system is achieved through two AWGs that synchronously drive the MZM, PM, AOM, and variable BS. The shorter loop integrates both PM and MZM for dynamic phase shifter and gain/loss modulations, with the MZM's gain/loss factors limited to $[e^{-0.3}, e^{0.3}]$.

Each loop contains an erbium-doped fiber amplifier (EDFA) to compensate for MZM-induced suppression loss and passive damping effects. Gain stabilization in the EDFA is achieved through pre-injection of 1530 nm pilot light, which partially saturates the amplifier medium to maintain energy consistency. Amplified spontaneous emission noise is suppressed by an OTF positioned downstream of the EDFA. Polarization stability is enforced via a polarization BS that filters orthogonal polarization components, while isolators ensure unidirectional propagation and polarization controllers (PCs) provide fine polarization adjustments. For signal monitoring, 50:50 BS routes portions of the light to photodetector, with the converted electrical signals captured by a digital storage oscilloscope (DSO).

The time-synthetic ONNs autonomously interact with hardware equipment through a VISA interface: optical intensities from the DSO are streamed to GPUs for real-time analysis, while optimized control voltages for PM/MZM modulation are transmitted to AWG. During the training phase, the time-synthetic ONNs dynamically adjust phase shifters and gain/loss factors in response to time-varying noise, achieving adaptive control through continuous feedback. The full automation of the experimental platform enables in-situ training without human intervention, hence eliminating the dataset bias and optimizing the operational efficiency.

### Propagation operator expression

The propagation operators of the time-synthetic lattice function as weight matrices of ONNs, derived from the discrete quantum walk equations. Due to spatial asymmetry, their expressions take two forms. For even time layers:

$$W^{2n} = \begin{bmatrix} H_i^{2n} & 0 & 0 & 0 & 0 & 0 \\ 0 & H_{l,1}^{2n} & 0 & 0 & 0 & 0 \\ 0 & 0 & H_{l,2}^{2n} & 0 & 0 & 0 \\ 0 & 0 & 0 & \ddots & 0 & 0 \\ 0 & 0 & 0 & 0 & H_{l,M\text{-}1}^{2n} & 0 \\ 0 & 0 & 0 & 0 & 0 & H_{r,M}^{2n} \end{bmatrix} \tag{1}$$

where the block submatrices can be expressed as follows:

$$H_i^{2n} = \begin{bmatrix} 0 & 0 \\ i\sqrt{2}/2 & 0 \end{bmatrix} \tag{2}$$

$$H_{l,m}^{2n} = \begin{bmatrix} 0 & i\sqrt{2}G_m^{2n}e^{i\varphi_m^{2n}}/2 & \sqrt{2}G_m^{2n}e^{i\varphi_m^{2n}}/2 & 0 \\ 0 & 0 & 0 & 0 \\ 0 & 0 & 0 & 0 \\ 0 & \sqrt{2}/2 & \mathrm{i}\sqrt{2}/2 & 0 \end{bmatrix} \tag{3}$$

$$H_{r,M}^{2n} = \begin{bmatrix} 0 & i\sqrt{2}G_M^{2n}e^{i\varphi_M^{2n}}/2 \\ 0 & 0 \end{bmatrix} \tag{4}$$

where the superscript $2n$ denotes the time layer, and the subscript $m$ represents the position.

For odd time layers:

$$W^{2n+1} = \begin{bmatrix} H_{l,1}^{2n+1} & 0 & 0 & 0 \\ 0 & H_{l,2}^{2n+1} & 0 & 0 \\ 0 & 0 & \ddots & 0 \\ 0 & 0 & 0 & H_{l,M}^{2n+1} \end{bmatrix} \tag{5}$$

where the block submatrices $H_{l,m}^{2n+1}$ retain the same form as Eq. 3. Unlike the propagation operator in other related works[21], which spans two round trips, in this work, each propagation operator corresponds to one round trip to facilitate the measurement of in-situ backpropagation gradients. The expression becomes equivalent to that of the previous one by multiplying the operators of two consecutive time layers.

Analogous to the spatial Clements design[45], the time-synthetic ONNs achieve full connectivity over *N* round trips, where *N* corresponds to the number of input channels. By dynamically modulating the beam-splitting ratio, gain/loss, and phase shift of each time gate, the fully-connected time-synthetic ONNs can realize arbitrary *N* × *N* complex-valued linear transformations.

**Gain/loss gradient measurement for in-situ training**

To accelerate training speed and enhance accuracy, gain/loss gradients are decomposed via the chain into several items that can be directly measured during forward propagation. For the output layer, the gradient expression is: $\frac{\partial \mathcal{L}}{\partial G_\theta^n} = \mathcal{R}\left\{(\widehat{Y^{out}} - Y)^\dagger \frac{\partial W^n}{\partial G_\theta^n}\widehat{Y^n}\right\}$, and error propagation follows: $\frac{\partial \mathcal{L}}{\partial \widehat{Y^n}} =$ $(W^n)^{\mathrm{T}}(\widehat{Y^{out}} - Y)^{\dagger}$. For hidden layers, the gradient expression is: $\frac{\partial \mathcal{L}}{\partial G_{\theta}^{n\text{-}1}} = \mathcal{R}\left\{\frac{\partial \mathcal{L}}{\partial \widehat{Y^{n}}} \frac{\partial W^{n\text{-}1}}{\partial G_{\theta}^{n\text{-}1}} \widehat{Y^{n\text{-}1}}\right\}$, and error propagation follows: $\frac{\partial \mathcal{L}}{\partial \widehat{Y^{n\text{-}1}}} = (W^{n\text{-}1})^{\mathrm{T}} \frac{\partial \mathcal{L}}{\partial \widehat{Y^{n}}}$.

In these expressions, $\widehat{Y^{n\text{-}1}}$, $\widehat{Y^{n}}$ and $\widehat{Y^{out}}$ denote detected optical intensities in different time layers, while $Y$ represents one-hot encoded target output (a given item). Activation values of artificial neurons in adjacent time layers satisfy: $\widehat{Y^{n}} = W^{n\text{-}1}\widehat{Y^{n\text{-}1}}$, allowing the weight matrix to be measured as $W^{n\text{-}1} = \widehat{Y^{n}}(\widehat{Y^{n\text{-}1}})^{\text{-}1}$.

With phase modulation excluded during in-situ training, Eqs. 3 and 4 can be further simplified to:

$$H_{l,m}^{2n} = \begin{bmatrix} 0 & i\sqrt{2}G_m^{2n}/2 & \sqrt{2}G_m^{2n}/2 & 0 \\ 0 & 0 & 0 & 0 \\ 0 & 0 & 0 & 0 \\ 0 & \sqrt{2}/2 & i\sqrt{2}/2 & 0 \end{bmatrix} \tag{6}$$

$$H_{r,M}^{2n} = \begin{bmatrix} 0 & i\sqrt{2}G_M^{2n}/2 \\ 0 & 0 \end{bmatrix} \tag{7}$$

Notably, the derivative matrix $\frac{\partial W^n}{\partial G_{\theta}^{n}}$ contains only constant elements: $0$, $i\sqrt{2}/2$ and $\sqrt{2}/2$. Consequently, all items of gain/loss gradients $\frac{\partial \mathcal{L}}{\partial G_{\theta}^{n}}$ are either experimentally measurable, constant, or given. In this way, the time-synthetic ONNs can dynamically adapt to the time-varying errors in real-world environments.

**Achieving a nonlinear activation function in a linear system**

We simulate a time-synthetic lattice propagating through 40 round trips with a beam-splitting ratio of $\beta = \pi/4$. To validate the first type of nonlinearity, a phase shifter $\varphi \in [0, \pi]$ is modulated at a designated lattice site (Extended Data Fig. 2a, purple). For the second type, a gain/loss factor $G \in [e^{-0.3}, e^{0.3}]$ is modulated at two lattice sites (Extended Data Fig. 2a, green), consistent with the input encoding schemes in this work. Then we calculate the output intensities at four positions in the shorter loop (Extended Data Fig. 2b and 2c), and the nonlinear relationship between input ($G$ and $\varphi$) and output intensities clearly demonstrates the effectiveness of the structural nonlinearity.

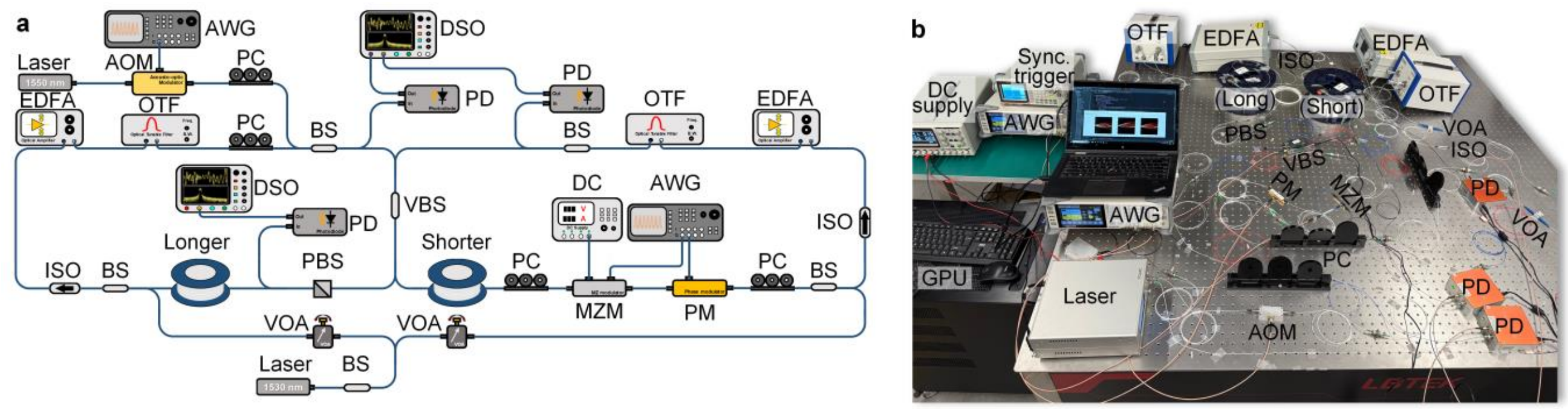

**Extended Data Fig. 1 | a** Experimental pulse propagation path diagram. **b** Real shot of experimental setup. AWG: arbitrary waveform generator; DSO: digital storage oscilloscope; AOM: acousto-optic modulator; PC: polarization controller; BS: beam splitter; VBS: variable beam splitter; PBS: polarizing beam splitter; PD: photodetector; EDFA: erbium-doped fiber amplifier; OTF: optical tunable filter; ISO: isolator; MZM: Mach-Zehnder modulator; PM: phase modulator; DC: direct current; VOA: variable optical attenuator.

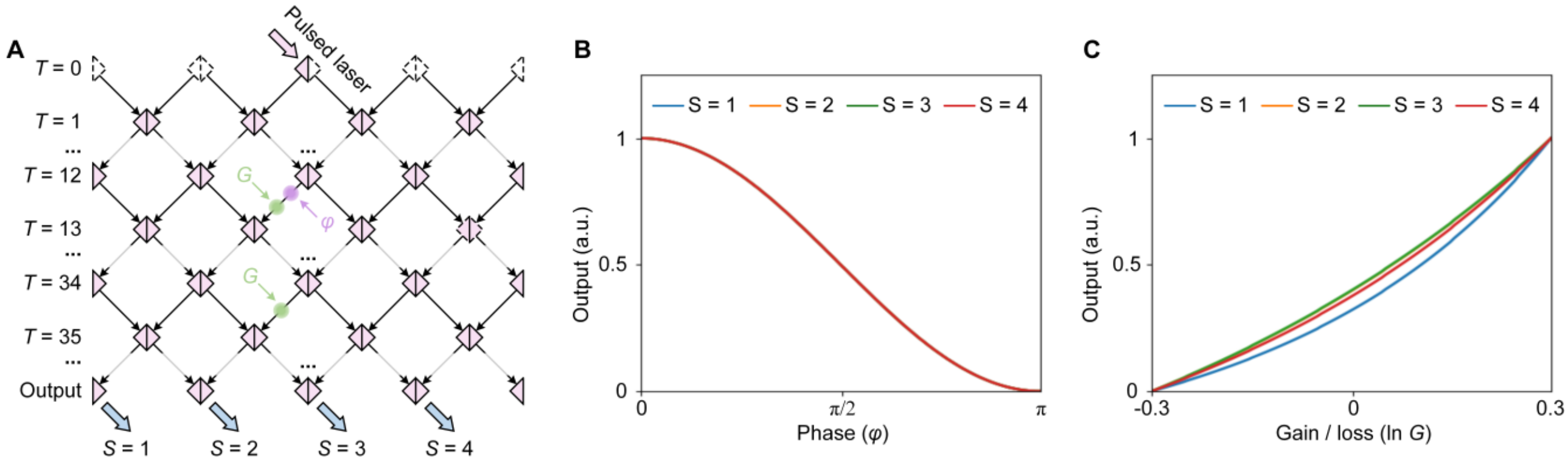

**Extended Data Fig. 2 | a** A time-synthetic lattice propagating through 40 round trips. **b**-**c** Output intensities at four positions in the shorter loop as a function of (**b**) the phase shifter and (**c**) the gain/loss factor, demonstrating the nonlinear input-output relationship.